\documentclass[aps,twocolumn]{revtex4}
\usepackage[dvips]{color}
\usepackage{epsfig}

\usepackage[novolumeabbr]{unitsdef} 
\newunit{\invcm}{\centi\meter\unitsuperscript{-1}}

\begin{document}

\title{Transparent and flexible polymerized graphite oxide thin film with frequency-dependent dielectric constant}

\author{D.~W.~Lee$^{1*}$, J.~W.~Seo$^{1,2}$, G.~R.~Jelbert$^{1,2}$, L.~de Los Santos V.$^{1}$, J.~M.~Cole$^{1,3}$, C.~Panagopoulos$^{1,2,4}$, and C.~H.~W.~Barnes$^{1}$}

\affiliation{$^{1}$ Cavendish Laboratory, University of Cambridge,
Cambridge CB3 0HE, United Kingdom}
\affiliation{$^{2}$ Division of Physics and Applied
Physics, Nanyang Technological University, 6373616 Singapore}
\affiliation{$^{3}$ Department of Chemistry and Physics,
University of New Brunswick, P. O. Box 4400, Fredericton, NB, E3B5A3, Canada}
\affiliation{$^{4}$ Department
of Physics, University of Crete and FORTH, 71003 Heraklion,
Greece}

\begin{abstract}
Here we report on the preparation of transparent and flexible polymerized graphite oxide, which is composed of carbons with $sp^{3}$-hybridized orbitals and a non-planar ring structure, and which demonstrates dispersion in its dielectric constant at room temperature. This frequency dependence renders the material suitable for creating miniaturized, flexible, and transparent variable capacitors, allowing for smaller and simpler integrated electronic devices. We discuss this polarizability in terms of space charge effects.
\end{abstract}

\maketitle
The development of micro- and nano-electronics enabled the revolution of modern devices. Much effort has gone into the synthesis and understanding of non-volatile ferroelectric materials, as they hold the key for overcoming the ageing problems of devices such as ferroelectric random access memory (FRAM)\cite{Park,Junquera,DWLee}. Dielectrics have also been the subject of extensive study. The system with the lowest dielectric constant determines miniaturizability\cite{Reynard}, as it separates the conducting wires and the transistors. On the other hand, gate insulators need a high dielectric constant to reduce leakage current and therefore power consumption\cite{Kaushik}. However, there has been limited research on materials with variable dielectric constants, despite the large potential impact a robust microvariable capacitor would have on microelectronics. Conventional variable capacitors are tuned with mechanical or electrical volume changes and are therefore relatively large.

An attractive method for creating materials with variable dielectric constants is to exploit the phenomenon of space charge. Space charge is usually induced in insulating materials with a layered structure. Graphite, though layered [Fig. 1(a)], is conductive. It is composed of carbon atoms with $sp^{2}$-$sp^{2}$ bonding, different from allotropes like diamond which are made up of sp3-hybridized carbon. To increase an amount of space charge in graphite, it must lose its conductivity by oxidation and chemical groups must be formed on its surface. If graphite is oxidized in air, it decomposes into carbon dioxide. However, with carefully controlled synthesis, it can be oxidized into an insulating, dark brown graphite oxide\cite{Brodie,Lerf3}. Although the precise structure is debatable\cite{Lerf3,12}, it is generally agreed that graphite oxide has epoxy (-O-) and hydroxyl (-OH) groups and it consists of $sp^{2}$- and some $sp^{3}$-hybridized carbon [Fig. 1(b)] which causes distortions in the layers. Both the epoxy and hydroxyl groups and the distortions can induce space charges between the layers. More space charges are induced between the layers after epoxy groups are decomposed in graphite oxide [Fig. 1(c)]. Since carbons bonded with -OH and -ONa groups have $sp^{3}$-hybridized orbitals, the change of the orbitals from $sp^{2}$ to $sp^{3}$ causes the distortion and leads the sample to have a nonaromatic ring structure like cyclohexane [Fig. 1(d)]. Figure 1(e) shows the schematic structure of the sample which is made up of nonaromatic rings of carbons and is totally different from diamond despite the fact that both of them consist of carbons with $sp^{3}$-hybridized orbitals.

In this experiment, graphite oxide samples were prepared by the Brodie process, in which 5.0\gram of graphite is added to 62.5\milliliter of fuming nitric acid. After cooling the mixture in an ice bath, 25.0\gram of potassium chlorate is added slowly. After the mixture reaches room temperature, it is placed in a water bath, heated slowly to 45\celsius and kept at this temperature for 20\hour. The mixture is then poured into 125\milliliter of cold distilled water and warmed to 70\celsius, centrifuged, and decanted. Graphite oxide samples were obtained by being washed three times in this manner and dried overnight at 70\celsius.

\begin{figure}[!b]
\epsfig{file=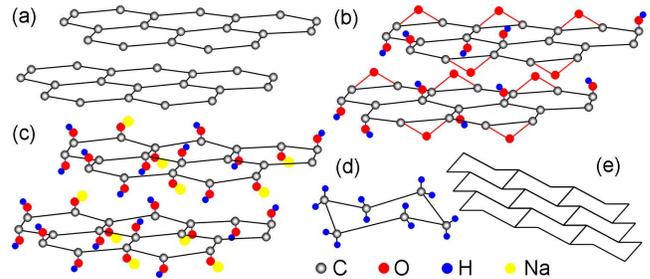, width=8.5cm}
\caption{(Color online) Schematic structure of (a) graphite (b) graphite oxide (c) NaOH-reacted graphite oxide (d) cyclohexane and (e) polymerized graphite oxide.}
\end{figure}

Epoxy groups in graphite oxides are vulnerable to NaOH, decomposing into hydroxyl groups (-OH) and sodium oxide groups (-ONa). So, if graphite oxide is immersed in a NaOH solution, most epoxy groups are destroyed and it turns black. It starts to become flexible after 1 week when the polymerization is thought to occur. After 2 weeks, flexible graphite oxide film was obtained after being washed with de-ionized (DI) water and dried in the air. The film is flexible and plasticlike. It differs from powders in that the film is wholly connected. For these reasons, we believe it is a polymer although the mechanism of the polymerization is not clear. The thickness of the film can be controlled by changing the concentration of graphite oxide in the NaOH solution. For example, if 0.01\gram of graphite oxide is immersed in a NaOH solution and is transferred to 25\milliliter of DI water after being washed with the same, it forms a substrate approximately 1\micrometer thick. However, if 0.1\gram of graphite oxide is transferred to 5\milliliter of DI water, the thickness of the film is ca. 0.1\millimeter.

\begin{figure}[!t]
\epsfig{file=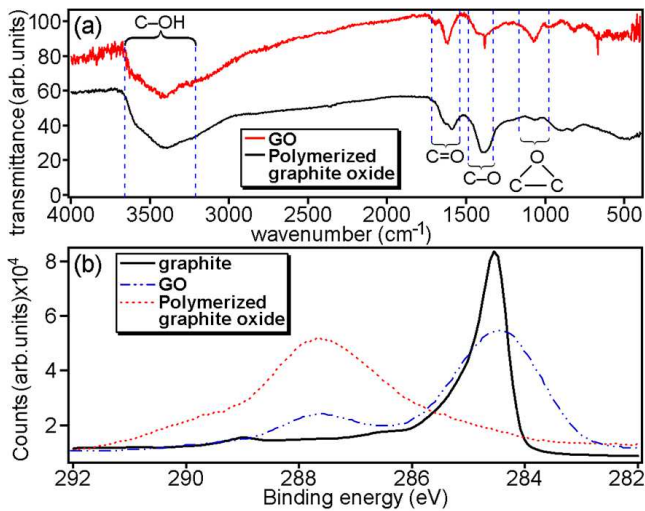, width=8.5cm}
\caption{(Color online) Characterization of thin-polymerized graphite oxide using FT-IR and XPS. (a) The FT-IR spectrum from graphite oxide and polymerized graphite oxide. The epoxy group decomposes during polymerization. (b) Carbon core-level XPS spectra of graphite, graphite oxide, and polymerized graphite.}
\end{figure}

To characterize the structure of the samples, we used a Thermo Nicolet Avatar 360 Fourier transform infrared (FT-IR) spectrometer. Figure 2(a) shows the FT-IR spectra of graphite oxide and polymerized graphite oxide. Four main peaks have been identified, centered at 1050, 1380, 1650, and 3470\invcm. The peak at 1050\invcm represents epoxy groups while the presence of hydroxyl groups is confirmed by the broad peak at 3470\invcm, denoting C-OH stretching. The peak at 1380\invcm corresponds to a C.O vibrational mode whereas that at 1650\invcm represents ketone groups which have localized $\pi$ electrons. It can be seen that most epoxy groups decompose in reaction with NaOH and the peak intensity at 1380\invcm increases as a result of break-up of epoxy groups to hydroxyl groups.

To investigate the electronic structure of the samples, x-ray photoemission spectroscopy (XPS) experiments were performed on the BACH beamline at Elettra in Italy. Figure 2(b) illustrates core-level carbon XPS data of graphite (99.99+\% purity and 45\micrometer, Aldrich), graphite oxide, and polymerized graphite oxide. The peak centered at 284.6\eV is assigned to carbon-carbon (C-C) bonds in aromatic networks in graphite and has a well-known asymmetric line shape\cite{Javier}. Graphite oxide has two peaks at 285 and 287.6\eV. The peak at 285\eV is from carbon atoms with $sp^{2}$-hybridized orbitals. The peak at approximately 287.6\eV originates from C-O which is bonded with $sp^{3}$-hybridized orbitals\cite{Cheung,Katzman,Desimoni,Papirer,Hiroki}. Unlike graphite, most carbon atoms in polymerized graphite oxide have $sp^{3}$-hybridized orbitals similar to diamond even though some carbon atoms in the sample have $sp^{2}$-hybridized orbitals.

\begin{figure}[!t]
\epsfig{file=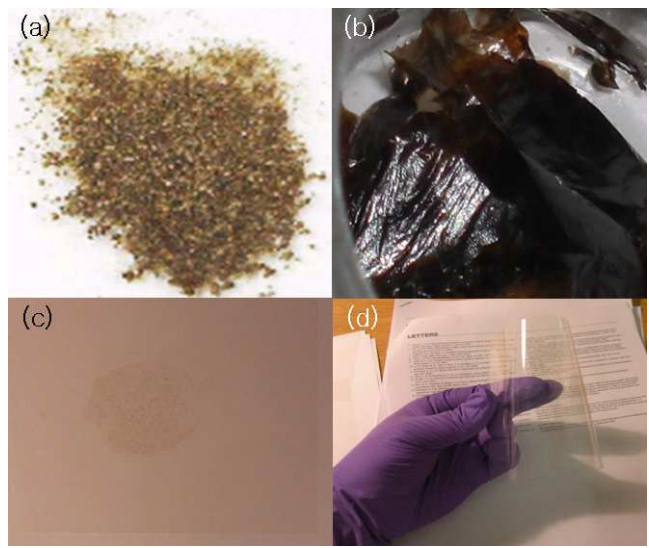, width=8.5cm}
\caption{(Color online) Sample pictures of graphite oxide. (a) NaOH-reacted graphite oxide; (b) thick-polymerized graphite oxide; (c) thin-polymerized graphite oxide on a transparent substrate; and (d) thin-polymerized graphite oxide on a flexible substrate.}
\end{figure}

Figure 3 compares graphite oxide, thick-polymerized graphite oxide and thin-polymerized graphite oxide. Graphite oxide is brown in color and is an insulator (FIG.~3(a)). Thick-polymerized graphite oxide is flexible and appears black (FIG.~3(b)) while thin-polymerized graphite oxide is transparent (FIG.~3(c)) and flexible (FIG.~3(d)). Furthermore, it can be grown on plastic substrates and transferred with ease, making it a potential candidate for manufacturing flexible electronic and optical devices.

The thickness of the polymerized graphite oxide on the transparent flexible plastic substrate shown in Figs. FIG.~3(c)-(d) was measured with a VEECO Dimension 3100 atomic force microscope (AFM) in tapping mode. Figure 4(left) shows the sample and substrate on the AFM sample stage. Since both the substrate and the sample are almost transparent, most of the shapes and contrasts shown are due to the sample stage. The left of each picture shows the substrate, whereas the right includes the thin-polymerized graphite oxide. The thickness scale in the image (FIG.~4 (top right)) is 1.2\micrometer. The substrate is flat and the edge between the sample and the substrate is clearly visible. The scanning profile of the white line in the AFM image is shown in FIG.~4 (bottom right), revealing a gradient at the edge of the sample. The thickness of the sample is approximately 1\micrometer.

\begin{figure}[!t]
\epsfig{file=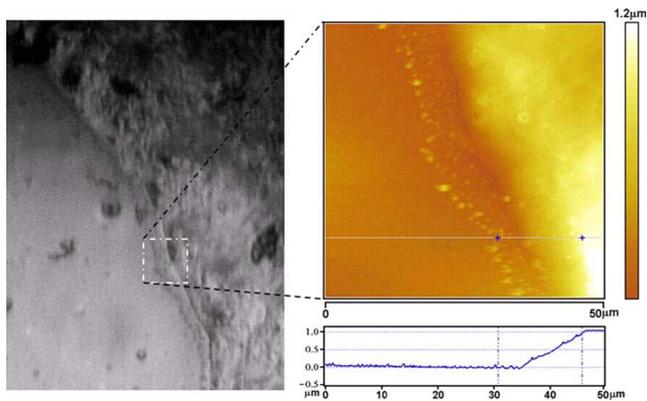, width=8.5cm}
\caption{(Color online) Characterization of thin-polymerized graphite oxide using AFM. Sample pictures of thin-polymerized graphite oxide on a transparent flexible plastic substrate on the AFM sample stage with an optical camera (left) and an AFM image (right top). The thickness profile of the sample (bottom right) was measured along the white line shown in the AFM image.}
\end{figure}

To elucidate the low frequency charge dynamics, we performed low frequency dielectric spectroscopy. The sample was prepared with silver paste contacts on either side in a capacitative configuration, the impedance $Z$ and phase angle $\theta$ were measured with an Agilent 4284 A precision LCR meter, and the ac voltage of 0.1\volt was applied to the sample of which thickness was 0.014 cm. Figure 5(a) depicts the real part of the dielectric constant for the polymerized graphite oxide as a function of temperature for different frequencies (from 77 to 500\kilohertz). When the frequency is 77\hertz, the value of the dielectric constant is approximately 230 at 293\kelvin. It decreases rapidly as the temperature decreases for a given frequency. Below $\sim$250\kelvin, the dielectric constant is less than 10 and is independent of temperature. The value of the dielectric constant decreases with increasing frequency. The frequency-dependent behavior of the dielectric constant is depicted in Fig. 5(b). At T=292\kelvin, the value of the dielectric constant is approximately 200 for the lowest frequency. At a given frequency, the value of the dielectric constant decreases as the temperature decreases. With the temperature fixed, the value rapidly decreases as the frequency increases and becomes constant above $\sim$10\kilohertz. The dissipation factor ($\tan \delta$ = $\epsilon_{i}/\epsilon_{r}$, where $\epsilon_{r}$ and $\epsilon_{i}$ are the real and imaginary parts of the dielectric constant, respectively) is shown in Fig. 5(c). The peak of the dissipation factor appears around 273\kelvin at lower frequency and is shifted toward higher temperature as frequency increases. To examine the electrical properties of the sample in more detail, we extracted the resistivity as a function of frequency and temperature, as shown in Fig. 5(d). The resistance in Fig. 5(d) increases with temperature until it reaches a maximum and then it decreases again. The peak is shifted toward higher temperatures as the frequency is increased. This behavior
is similar to that of positive temperature coefficient resistance (PTCR) materials\cite{HEYWANG1,HEYWANG2,Jonker1}. At low electric field the dielectric constant of PTCR materials increases as temperature increases, and vice versa at high electric fields\cite{HEYWANG1,HEYWANG2}. Heywang\cite{HEYWANG1,HEYWANG2} and Jonker\cite{Jonker1,Jonker2} suggested that space charges are involved in the properties of the dielectric constant as a function of temperature in PTCR materials. We also can understand the temperature and frequency dependence of the dielectric constant of the samples in terms of the space charge generated by charge-transfer from the carbon atoms to the oxygen atoms. Charge-transfer mainly depends on the
electronegativity of the atoms and the structure of the material. Since oxygen atoms have higher electronegativity than carbon, hydrogen, and sodium (the electronegativity of carbon is 2.55, that of sodium is 0.93, that of hydrogen is 2.20, and that of oxygen 3.44\cite{electronegativity}, sodium and hydrogen bonded with oxygen has positive charge (Na$^{+}$ and H$^{+}$) while oxygen has negative charge (O$^{2-}$). As more hydroxyl groups and .ONa groups are produced between the layers of the sample, more charges are transferred to oxygen atoms and more charges are localized between the layers. As a result, the sample has a large number of space charges which we believe cause the dielectric dispersion seen\cite{Tanaka1}.

\begin{figure}[!t]
\epsfig{file=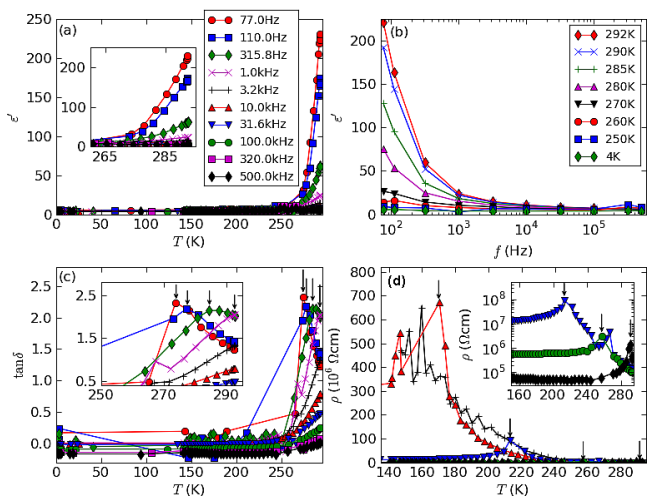, width=8.5cm}
\caption{(Color online) Electrical properties of thin-polymerized graphite oxide, showing (a) the temperature-dependence, (b) the frequencydependence of the dielectric constant, (c) the dissipation factor, and (d) the resistivity. The insets show enlargements of the higher temperature regions. The lines are guides to the eye. The arrows show the peaks discussed in the text.}
\end{figure}

In summary, we demonstrate the potentiality of room temperature dielectric devices from flexible and transparent polymerized graphite oxide. We propose that space charges generated on sodium and hydrogen atoms bonded with oxygen are the origin of the high dielectric constant at room temperature. The novel properties such as flexibility, transparency, and dispersion in its dielectric constant at room temperature show the enormous potential for replacing conventional (large volume) variable capacitors and for incorporation into mobile devices or flexible displays.

\section*{Acknowledgements}
D. W. Lee is grateful to L. M. Brown for helpful discussions. L. de Los Santos V. thanks the European Union Programme for Latin America, ALBAN (Grant No E06D101257PE) and Cambridge Overseas Trust for financial support. J. M. Cole acknowledges support from The Royal Society and UNB Vice-Chancellor's research chair. The work in Crete and Singapore was supported by the MEXT-CT-2006-039047, EURYI, and the National Research Foundation of Singapore. This paper is written in memory of J. A. C. Bland, who suddenly passed away on 2 December 2007.

\clearpage

\end{document}